\documentclass[twocolumn,aps,superscriptaddress,preprintnumbers]{revtex4-1}

\usepackage{amsmath,amssymb}
\usepackage{graphicx}
\usepackage{dcolumn} 
\usepackage{bm} 
\usepackage{soul}
\usepackage{multirow}
\usepackage{setspace}
\usepackage[mathlines]{lineno}
\usepackage{textcomp}
\usepackage{mathcomp}
\usepackage{spverbatim}
\usepackage{float}
\usepackage{wasysym}
\usepackage{array}
\usepackage{fancyhdr}
\usepackage{subfigure}
\usepackage{physics}
\usepackage{xcolor}

\usepackage{hyperref}
\hypersetup{
    colorlinks=true,       
    linkcolor=blue,        
    citecolor=blue,        
    filecolor=blue,        
    urlcolor=blue,         
    runcolor=blue
}

\usepackage{enumitem,amssymb}
\newlist{todolist}{itemize}{2}
\setlist[todolist]{label=$\square$}
\usepackage{pifont}

\newcommand{\USACHP}{Department of Physics, Universidad de Santiago de Chile, Av. Victor Jara 3493, Santiago, Chile}
\newcommand{\UNDC}{Department of Chemical and Biomolecular Engineering, University of Notre Dame, IN, USA}
\newcommand{\USACHC}{Department of Chemistry of Materials, Universidad de Santiago de Chile, Santiago, Chile.}
\newcommand{\MIRO}{Millennium Institute for Research in Optics, Concepción, Chile}

\begin{document}
\renewcommand{\figurename}{Figure}
\renewcommand{\tablename}{Table}

\title{Semi-Empirical Haken-Strobl Model for Molecular Spin Qubits}

\author{Katy Aruachan}
\affiliation{\USACHP}
\author{Yamil J. Colón}
\affiliation{\UNDC}
\author{Daniel Aravena}
\affiliation{\USACHC}
\author{Felipe Herrera}
\email{felipe.herrera.u@usach.cl}
\affiliation{\USACHP}
\affiliation{\MIRO}

\date{\today}

\begin{abstract}
Understanding the physical processes that determine the relaxation $T_{1}$ and dephasing $T_2$ times of molecular spin qubits is critical for envisioned applications in quantum metrology and information processing. Recent spin-echo $T_1$ measurements  of solid-state molecular spin qubits have stimulated the development of quantum mechanical models for predicting  intrinsic spin qubit timescales using first-principles electronic structure methods. We develop an alternative semi-empirical approach to construct Redfield quantum master equations for molecular spin qubits using a stochastic Haken-Strobl model for a central spin with a fluctuating gyromagnetic tensor due to spin-lattice interaction and a fluctuating local magnetic field due to interactions with other lattice spins. Using a vanadium-based spin qubit as a case study, we compute qubit population and decoherence timescales as a function of temperature and magnetic field using a bath spectral density parametrized with a small number of $T_{1}$ measurements. The theory quantitatively agrees with experimental data over a range of conditions beyond those used to parametrize the model, demonstrating the generalization potential of the method. The ability of the model to describe the temperature dependence of the ratio $T_2/T_1$ is discussed and possible applications  for designing novel molecule-based quantum magnetometers are suggested. 

\end{abstract}
\maketitle
\section{\label{sec:intro}Introduction}
Electron spins in molecules have emerged as promising qubit systems \cite{atzori2016room,atzori2016quantum, yu2019concentrated, gaita2019molecular,Wasielewski:2020tp,Tesi2016} due to their relatively long intrinsic relaxation and decoherence timescales, comparable with traditional solid-state spin systems such as NV-centers \cite{zadrozny2015millisecond,graham2017forging, awschalom2018quantum}. Molecular spin qubit architectures also benefit from the versatility enabled by nanoscale integration with other solid-state platforms \cite{Jenkins_2013,Ghirri2016,Bonizzoni2016,Bonizzoni2018,Gimeno:2020,Chiesa2023} and the potential for manipulating the local qubit environment using synthetic chemistry techniques \cite{gaita2016coherence,yamabayashi2018scaling,Amdur2022}. Future advances in spin qubit coherence and controllability would require a precise characterization of the mechanisms that lead to spin decoherence in molecular materials \cite{Lavroff:2021ww}. 

Multi-scale {\it ab-initio} modeling techniques have been proposed and successfully used to predict population $(T_1)$ and decoherence $(T_2)$ timescales of solid-state molecular spin qubits, only using the chemical composition and geometry of the underlying crystal lattice as input \cite{albino2019first,Lunghi2019, lunghi2022toward, lunghi2017role, lunghi2017intra, lunghi2022computational,bayliss2022enhancing}. First-principles techniques offer valuable atomistic insights that can be used for designing chemical strategies aimed to improve the spin qubit performance. However,  the large computational overhead of atomistic simulations is a challenge for implementatiion of large-scale computational discovery strategies for molecular spin qubit materials.  

 \begin{figure}[t]
    \centering
    \includegraphics[width=0.5\textwidth]{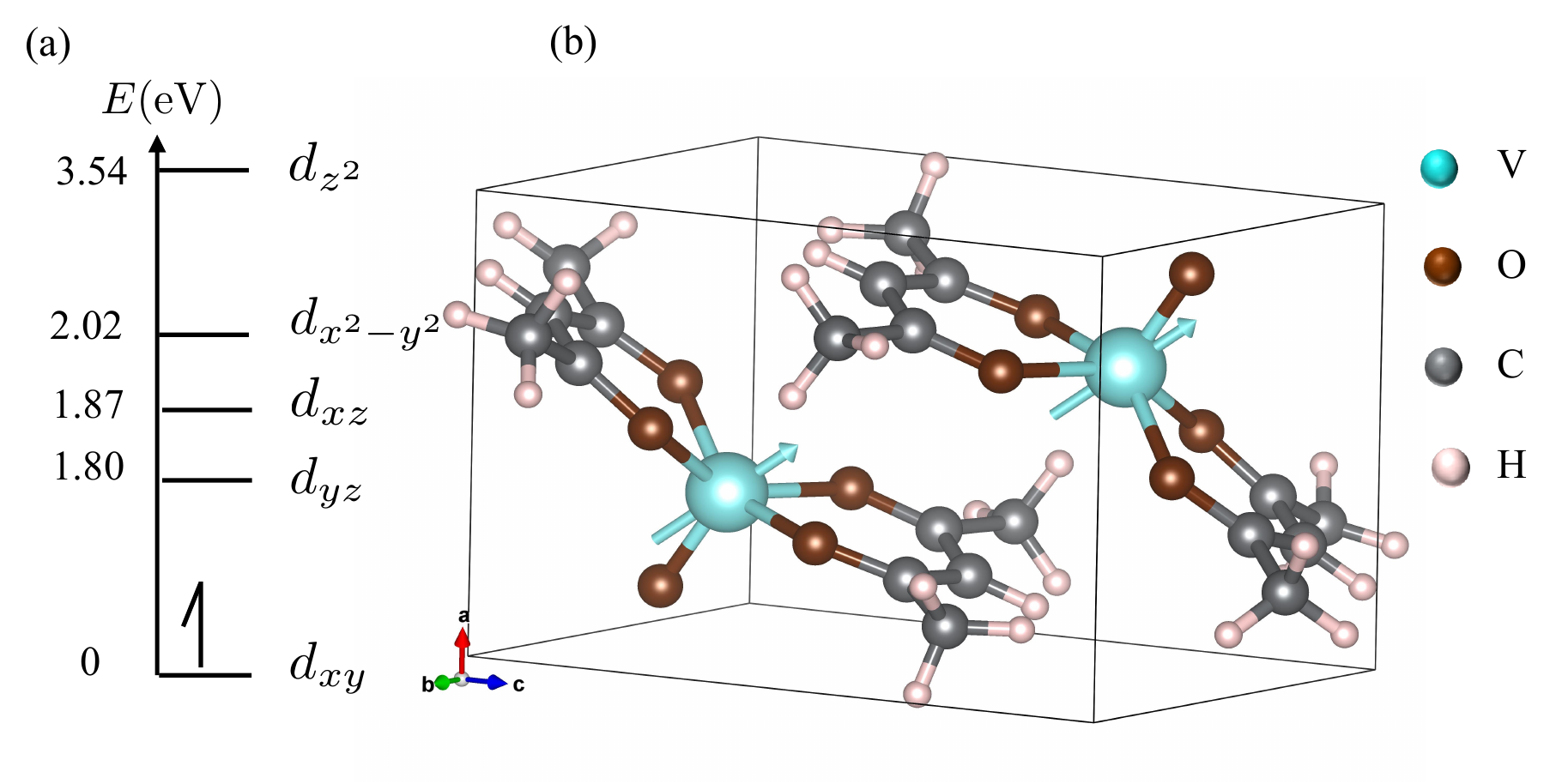}
    \caption{{\bf Vanadium(IV) molecular spin qubit}. (a) 3$d$-orbital splitting of VO$^{2+}$ in the ligand field of acetylacetonate (acac). (b) Unit cell of $\mathrm{VO(acac)_{2}}$, highlighting vanadium (cyan), oxygen (brown), carbon (gray), and hydrogen atoms (pink).  The central spin angular momentum $\vec{S}$ in the V$^{\mathrm{IV}}$ ion is represented by arrows at the metal centers.} 
    \label{fig:qubit-structure}
\end{figure}

Recently, the relaxation dynamics of high-spin single-molecule magnets was theoretically studied with a semi-empirical approach \cite{garlatti2021cost}. The dynamics problem was reduced in complexity by partitioning the crystal Hamiltonian in a part that is straightforward to compute using quantum chemistry packages and a parametrized term that models the interaction of the central spin with lattice phonons. The free model parameters were then determined by comparing theoretical predictions with magnetic relaxation measurements.  

In this work, we further develop this semi-empirical approach by constructing parametrized Redfield quantum master equations to describe the interaction of molecular spin qubits with lattice phonons and electron spin baths. The method is based on Haken-Strobl theory \cite{haken1972coupled,rips1993stochastic,chen2010effect}, which treats system-reservoir interactions as stochastic fluctuations of the system Hamiltonian. Haken-Strobl theory has been successfully used to study exciton transport and spectroscopy in molecular aggregates \cite{rips1993stochastic} and light-harvesting complexes \cite{ye2012excitonic}. We use it here to model spin-lattice interaction as a random fluctuation of the molecular gyromagnetic tensor of the spin qubit, as well as the interaction with a reservoir of electronic and nuclear spins embedded in the crystal lattice, which results in random fluctuations of the local magnetic field. 

Semi-empirical Redfield tensors can be used to compute relaxation and dephasing times, $T_1$ and $T_2$, for solid-state molecular spin qubits, by parametrizing the bath spectral density through a fitting procedure that matches the model prediction to a small number of $T_{1}$ measurements. We demonstrate the procedure  using experimental data for the vanadium(IV) complex illustrated in Fig. \ref{fig:qubit-structure}. This vanadium ion has an energetically isolated $3d_{xy}^1$ valence electronic configuration and thus represents an ideal $S=1/2$ spin system. The theory gives quantitatively accurate predictions for $T_{1}$ far beyond the experimental conditions used to parametrize the model, thus demonstrating the potential for generality of the method.



The rest of the article is organized as follows: in Section \ref{sec:TheFram} we construct the proposed semi-empirical quantum Redfield equation. In Section \ref{sec:ResAndDis} we discuss the $T_{1}$ and $T_{2}$ results for $\mathrm{VO(acac)_{2}}$ spin qubits as a function of temperature and magnetic field. In Section \ref{sec:ConAndOut} we conclude and discuss perspectives for  future work.

\section{\label{sec:TheFram}Theoretical Framework}
We use Haken-Strobl theory \cite{haken1972coupled,moix2013coherent} to describe the interaction of a central electron spin with magnetic and thermal environments. Following Ref. \cite{Reineker1975}, the total spin Hamiltonian is written as $\hat{H} (t)= \hat{H}_{S} + \hat{H}_{SB}(t)$, where $\hat{H}_{S}$ describes the stationary system and $\hat{H}_{SB}(t)$ describes the fluctuations of the system due system-bath interaction. To model the anisotropy of the electronic Zeeman interaction and capture the dependence of the spin dynamics on the magnetic field orientation \cite{abragam2012electron}, we write the total Hamiltonian as 
\begin{equation}\label{eq:total H}
\hat{H}(t)  = \frac{1}{\hbar}\mu_{B}(\vec{B} + \delta\vec{B}(t))\cdot (\overleftrightarrow{g} + \overleftrightarrow{\delta g}(t) )\cdot\vec{S}
\end{equation}
where $\mu_{B}$ is the Bohr magneton, $\hbar$ the reduced Planck constant, $\vec{B}$ the external magnetic field vector, $\stackrel{\leftrightarrow}{g}$ the gyromagnetic tensor, and $\vec{S}$  the spin angular momentum of the molecular qubit.  

We consider random fluctuations of the gyromagnetic tensor $\overleftrightarrow{\delta g}(t)$ as well as local magnetic field fluctuations $\delta B_{i}(t)$. The latter results from the presence of electronic and nuclear atomic spins the lattice structure. By neglecting higher order fluctuations, the total Hamiltonian can be partitioned into a central spin Zeeman term
\begin{equation}\label{E2}
 \hat{H}_{S}  = \frac{1}{2}\mu_{B}B_{i}g_{ij}\hat{\sigma}_{j},
 \end{equation}
where $g_{ij}$ are gyromagnetic tensor elements and $\hat{\sigma}_{j}$ are Pauli matrices, and a system-bath Hamiltonian of the form
\begin{equation}\label{E3}
 \hat{H}_{SB}(t)  =\left[ Q^{\delta g}_{j}(t) + Q^{\delta B}_{j}(t)\right] \hat{\sigma}_{j}  
 \end{equation}
where 
\begin{equation}\label{E4}
Q^{\delta g }_{j}(t) =\frac{1}{2}\sum_{i} \mu_{B}B_{i}\delta g_{ij}(t)
 \end{equation}
describes lattice-induced fluctuations of the $g$-tensor, and 
\begin{equation}\label{E5}
Q^{\delta B}_{j}(t) =\frac{1}{2}\sum_{i} \mu_{B} g_{ij} \delta B_{i}(t)
 \end{equation}
describe spin-bath fluctuations. Contraction over the Cartesian indices $i = (x,y,z)$ is implied throughout.

We focus on the dynamics of the reduced density matrix $\hat{\rho}_{s}(t)=Tr_{B}[\hat{\rho}(t)]$ in the Born approximation for the total state $\hat{\rho}(t) = \hat{\rho}_{s}(t)\otimes\hat{\rho}_{B}$, where $\hat{\rho}_{B}$ is the stationary bath state. In the Markov approximation, the vibrational and spin degrees of freedom of the reservoir relax faster than the system timescales. In the interaction picture with respect to $\hat{H}_{S}$, the  equation of motion for the matrix elements $\rho_{ab}^{s}$ of the reduced spin density matrix in the basis of system eigenstates $\ket{a}$, can generally be written as
\begin{equation}\label{eq:Redfield QME}
\frac{d}{dt}\rho_{ab}^{s}(t) = - \sum_{c,d}R_{ab,cd}\rho_{cd}^{s}(t)
\end{equation} 
where $R_{ab,cd}$ are elements of the Redfield tensor which encodes the relaxation and decoherence dynamics of the central spin system. The tensor elements are given by
\begin{eqnarray}\label{eq:Redfield tensor}
R_{ab,cd} &= & \frac{1}{\hbar^{2}}\left\{\delta_{bd}\sum_{e}\Gamma_{ae,ec}(\omega_{ce}) -  \right. \Gamma_{ca,bd}(\omega_{db}) \nonumber\\ 
&  & \left.- \Gamma_{db,ac}(\omega_{ca}) + \delta_{ac}\sum_{e}\Gamma_{be,ed}(\omega_{de}) \right\} 
\end{eqnarray}
with $\Gamma_{ab,cd}(\omega_{dc})$ being decay rate functions evaluated at the system transition frequencies $\omega_{ab} = \omega_{a} - \omega_{b}$. These rates can in turn be written in terms of spin selection rules and system-bath coupling strengths as
\begin{equation}\label{eq:Gamma}
\Gamma_{ab,cd}(\omega_{dc}) = {\rm Re}\left[\bra{a}\hat{\sigma}_{j}\ket{b}\bra{c}\hat{\sigma}_{j'}\ket{d}J_{jj'}(\omega_{dc})\right]. 
\end{equation} 
where 
$J_{jj'}(\omega)$ is the bath spectral density, which describes the coupling strength of the system with its environment at a given transition frequency and temperature \cite{deVega2017}. To model molecular spin qubits, we partition the total spectral density $J_{jj'}(\omega)$ in Eq. (\ref{eq:Gamma}) as  
\begin{equation}\label{E9}
J_{jj'}(\omega) = J_{jj'}^{\delta g}(\omega)  + J_{jj'}^{\delta B}(\omega), 
\end{equation}
here $J^{\delta g}_{jj'}(\omega)$ is the contribution from $g$-tensor fluctuations and $J^{\delta B}_{jj'}(\omega)$ describes field fluctuations due to couplings with the  spin bath.

\begin{table*}[t]
\centering
\begin{tabular}{| c | c | c | c | c | c | c | c | c |}
\hline  
Model&$a $ ($10^{-10}$ T$^2$)  & $c$ ($10^{-10}$) & $A$ ( $10^{-10}$ T$^2$)& $A_{g}$ ( $10^{-12}$ K$^{-\alpha}$) &  $\delta B$  ($\mu$T) &  $\delta g$ & $T_{2}$ HS Model (ms)   \\ \hline 
I   &0 & 0 & 0  & 4.0871 & 0  & 0.000403 &  2.149   \\
II  &5.5 & 0 & 5.5  & 10.823 &23.4& 0.000656 & 0.604   \\
III &5.5 & 1.5 &43 & 11.444 &65.5 &  $0.000674$ & 0.157 \\  \hline 
\end{tabular}
\caption{Spectral density parameters used in the $T_{1}$ and $T_{2}$ calculations in the text. The amplitude of the fluctuations of the local magnetic field are given by the polynomial $A(B) = a+ cB^{2}$. These sets of parameters are chosen to match the experimental measurement $T_{1}^{\rm exp}=0.940$ ms from Ref. \cite{Tesi2016}, for VO(acac)$_{2}$ at $B = 5$ T and $T = 10$ K. In all models the temperature scaling parameter is $\alpha = 4.6$} 
\label{tab:parameters}
\end{table*}

We construct $J^{\delta g}_{jj'}$ by writing the bath autocorrelation function as
\begin{equation}\label{eq:Q autocorrelation}
 \expval{Q^{\delta g}_{j}(\tau)Q^{\delta g}_{j'}(0)}_{B} = \left(\frac{\mu_{B}}{2}\right)^{2} B_{i}B_{i'}\expval{\delta g_{ij}(\tau)\delta g_{i' j'}(0)} 
\end{equation}
with $g$-tensor fluctuations of the form
\begin{equation}\label{eq:g fluctuation}
\expval{\delta g_{ij}(\tau)\delta g_{i' j'}(0)}  \equiv \sqrt{A_{ij}A_{i'j'}} T^{\alpha}e^{-\gamma_g \tau}\cos(\Omega_g \tau). 
\end{equation}
The amplitude matrix $A_{ij}$ is assumed to be isotropic for simplicity, i.e., $A_{ij}=A_{g}$, $\alpha$ is a temperature scaling power,  $\gamma_{g}$ and $\Omega_{g}$ give the bandwidth and the resonance frequency of the spectral density. These four parameters should be calibrated using experimental measurements of $T_{1}$ times, as explained below with an example material. Equation (\ref{eq:g fluctuation}) implies that the lattice-induced g-tensor fluctuations grows with temperature as a power law and have a finite correlation time $\tau_{g} \equiv 1/\gamma_{g}$ determined by the phonon spectrum, which is consistent with first-principles theory \cite{Lunghi2019}. 

The quantum mechanical spectral density $J_{jj'}^{\delta g}(\omega_{dc})$ that defines the Redfield tensor is constructed by taking the Fourier transform of the bath autocorrelation in Eq. (\ref{eq:Q autocorrelation}) multiplied by the semiclassical harmonic correction factor $\hbar\omega/2k_{B}T$, which is needed to satisfy detailed balance relations \cite{Egorov:1999vc,Valleau2012OnTA}. We obtain
\begin{eqnarray}\label{eq:J deltag}
J_{jj'}^{\delta g}(\omega) &= &  \frac{\hbar\omega}{2k_{B}T}\left(\frac{\mu_{B}}{2}\right)^{2} B_{i}B_{i'}\sqrt{A_{ij}A_{i'j'}}\,T^\alpha \nonumber\\
&&\times\left(\frac{1}{2\sqrt{2\pi}}\right)\frac{\gamma_{g}+i(\omega - \Omega_{g})}{\gamma_{g}^{2} + (\omega - \Omega_{g})^{2}} 
\end{eqnarray}
This expression vanishes at $\omega=0$ and therefore does not contribute to pure dephasing \cite{farrar2012pulse}. 

We follow a similar procedure to construct an effective spectral density that describes coupling to the low-frequency spin-bath $J_{jj'}^{\delta B}$, introducing the bath autocorrelation function
\begin{equation}\label{eq:QB autocorrelation}
\expval{Q^{\delta  B}_{j}(\tau)Q^{\delta B}_{j'}(0)} =  \left(\frac{\mu_{B}}{2}\right)^{2}g_{ij}g_{i'j'}\expval{\delta B_{i}(\tau)\delta B_{i'}(0)} 
\end{equation}
with the local magnetic field noise dynamics given by 
\begin{equation}\label{E14}
\expval{\delta B_{i}(\tau)\delta B_{i'}(0)}  \equiv A(B)\,e^{-\gamma_{\rm pd} \tau}.
\end{equation}
This expression describes pure dephasing with a finite amplitude at zero frequency given by
\begin{equation}\label{eq:AofB model}
A(B) = a+ c\,B^{2},
\end{equation}
where $B$ is the magnetic field magnitude. The quadratic field dependence of $A$ is compatible with a direct spin relaxation process \cite{abragam2012electron}.  The dephasing correlation time is $\tau_{\rm pd}=1/\gamma_{\rm pd}$. The parameters $a$ and $c$ determine the magnetic field dependence of the bath fluctuations and can be determined from $T_2$ measurements. Taking the Fourier transform of Eq. (\ref{eq:QB autocorrelation}), we obtain the spin bath spectral density
\begin{equation}\label{E16}
J_{jj'}^{\delta B}(\omega) =   \left(\frac{\mu_{B}}{2}\right)^{2} g_{ij}g_{i'j'}A(B)\left(\frac{1}{2\sqrt{2\pi}}\right)\frac{1}{\gamma_{\rm pd}  - i \omega}.
\end{equation}

To constrain the parameters $\{A_{g}, \alpha, \gamma_{g}, \Omega_{g}, a, c\}$ that determine the open quantum system model, we solve the Redfield equation for the reduced density matrix elements $\rho_{ab}^{s} = \bra{a}\hat{\rho}_{s}\ket{b}$ in the energy eigenbasis, using the total spectral density $J_{jj'}(\omega)$ from Eq. (\ref{E9}) with trial parameters. $T_{1}$ and $T_2$ times are obtained from the calculated decay dynamics of the populations $\rho_{aa}$ and coherences $\rho_{ab}$, and the parameters are iteratively improved by comparing the theoretical $T_{1}$ and $T_{2}$ times with suitable experimental measurements. Once the spectral density parameters are fixed to match a small set of measurements, the model predictions can be extrapolated over a broader set of conditions, as illustrated in what follows with a case study.

\section{Results and Discussion}
\label{sec:ResAndDis}

We demonstrate the parametrization of the spectral densities using selected measurements of $T_1$ times as a function of temperature and magnetic field for vanadyl acetylacetonate VO(acac)$\mathrm{_{2}}$ crystals, taken from Ref. \cite{Tesi2016}. For this molecular complex, we used the electronic structure methods described in Ref. \cite{ATANASOV2015,Ding2016} to obtain the diagonal $g$-tensor elements  $g_{xx} = 1.920102$, $g_{yy} = 1.978044$ and $g_{zz} = 1.981320$. The calculated qubit spectrum agrees well with measured Zeeman splittings for $B>0.2$ T.  The near-zero field spectrum is not well reproduced because explicit hyperfine interactions are ignored in Eq. (\ref{eq:total H}).

The spectral density parameters used to construct the Redfield tensor  are chosen to match the reported value $T_{1}^{\rm exp} = 0.940$ ms at $B = 5$ T and $T=10$ K, for the VO(acac)$\mathrm{_{2}}$ complex \cite{Tesi2016}. We constrain the parameters to match this measurement, obtaining $\Omega_{g} = 11$ cm$^{-1}$,  $\gamma_{g} = 2.387$ cm$^{-1}$, $\gamma_{\rm pd} = 0.1$ cm$^{-1}$, $A_{g} = 11.44\times 10^{-12}$ K$^{-\alpha}$,  $a= 5.5 \times 10^{-10}$ T$^{2}$, and $c= 1.5 \times 10^{-10}$. We note this is multi-valued parameter optimization problem that could be further constrained using multiple $T_1$ measurements. For the above magnetic field and temperature, the fitting procedure is most sensitive to the choice of $A_g$. The temperature scaling power $\alpha = 4.6$ is extracted from the measured temperature dependence of the $T_{1}$ time at $B = 5$ T, also from Ref. \cite{Tesi2016}. 

In Table \ref{tab:parameters}, we specify three sets of model parameters that give the same $T_{1}^{\rm exp} = 0.940$ ms. Model I assumes that only spin-lattice relaxation occurs. Models II assumes that the magnetic field fluctuations do not depend on the magnetic field magnitude and Model III assumes a quadratic  dependence with $B$. The magnitude of the bath-induced Hamiltonian fluctuations are defined as $\delta g \equiv \sqrt{\expval{\delta g_{ii}^{2}(0)}}$ and  $\delta B \equiv \sqrt{\expval{\delta B_{i}^{2}(0)}}$.

\subsection{Vanadyl Qubit Relaxation}

\begin{figure}[t]
   \centering
   \includegraphics[width=0.5\textwidth]{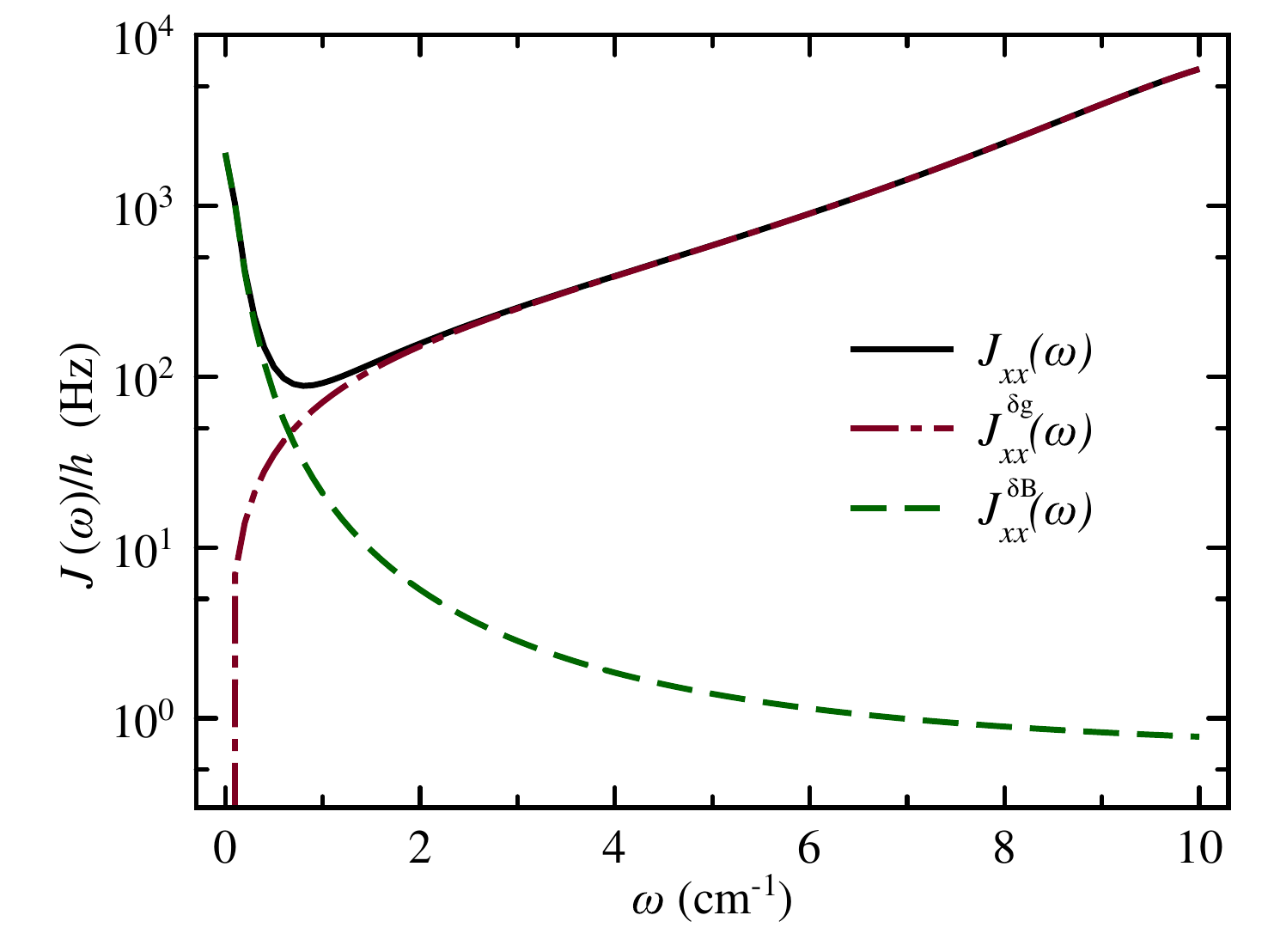}
   \caption{Total bath spectral density $J_{xx}(\omega)$ parametrized at $B = 5$ T and $T = 10$ K using experimental data from Ref. \cite{Tesi2016}. The  contributions of the phonon bath $J_{xx}^{\delta g}$ and spin bath $J_{xx}^{\delta B}$ are also shown. Spectral density parameters correspond to Model III in Table \ref{tab:parameters}.} 
   \label{fig:spectral density}
\end{figure}

Figure \ref{fig:spectral density} shows the spectral density $J_{xx}(\omega)$ obtained using the parameters from Model III in Table \ref{tab:parameters}. The contributions of $g$-tensor fluctuations $J_{xx}^{\delta g}$ and the spin bath $J_{xx}^{\delta B}$ are shown. At higher frequencies (large magnetic fields), $J_{jj'}^{\delta g} $ dominates the system-bath interaction by orders of magnitude, but its contribution to pure dephasing below 0.2 T is negligibly small. The finite zero-frequency amplitude of the spin bath spectral density ensures that the molecular spin qubit is coupled to the reservoir over the entire frequency range.

\begin{figure}[t]
\centering
\includegraphics[width=0.5\textwidth]{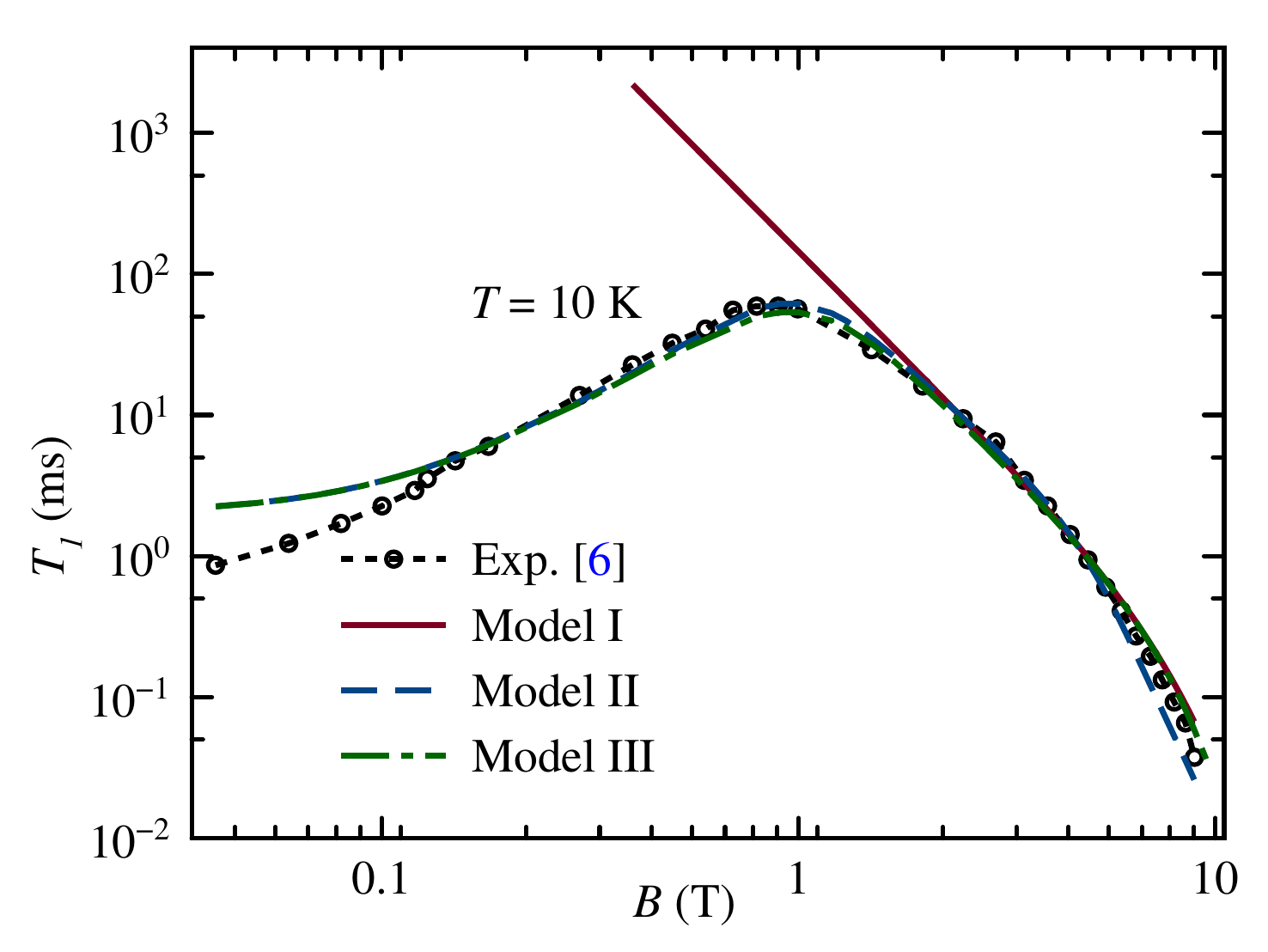}
\caption{Calculated spin relaxation time $T_{1}$ as a function of magnetic field $B$ using Haken-Strobl theory. Experimental $T_{1}$ data at $T = 10$ K are reproduced from Ref. \cite{Tesi2016} for comparison. Three spectral density models are compared (I-III), using parameters from Table \ref{tab:parameters}.} 
\label{F3}
\end{figure}

Figure \ref{F3} shows the calculated $T_{1}$ times for the VO(acac)$\mathrm{_{2}}$ spin qubit as a function of magnetic field at $T=10$ K. The predictions of the different models in Table \ref{tab:parameters} are compared with experiments. As expected and in agreement with ab-initio calculations \cite{Lunghi2019}, the reservoir model without magnetic field fluctuations (Model I) is in good agreement with $T_{1}^{\rm exp}$ at higher magnetic fields, where direct spin-phonon coupling dominates the relaxation dynamics. However, Model I is unable to capture the crossover around $B\approx 1\,{\rm T}$, below which  $T_1$ decreases with decreasing magnetic field, as the influence of the low-frequency spin reservoir becomes stronger. 

By including magnetic field fluctuations in the total spectral density, the theoretical predictions continue to reproduce the $T_1^{\rm exp}$ values  at higher magnetic fields, but now also capture the reduction of $T_1$ below the crossover. For $T=10\,{\rm K}$, we find that the using a field-dependent magnetic noise model (Model III) does not significantly modify the calculated crossover position and the $T_1$ values relative a theory with field-independent spin noise (Model II). 

\subsection{Effective Spin Dephasing Models}

\begin{figure}[t]
\includegraphics[width=0.5\textwidth]{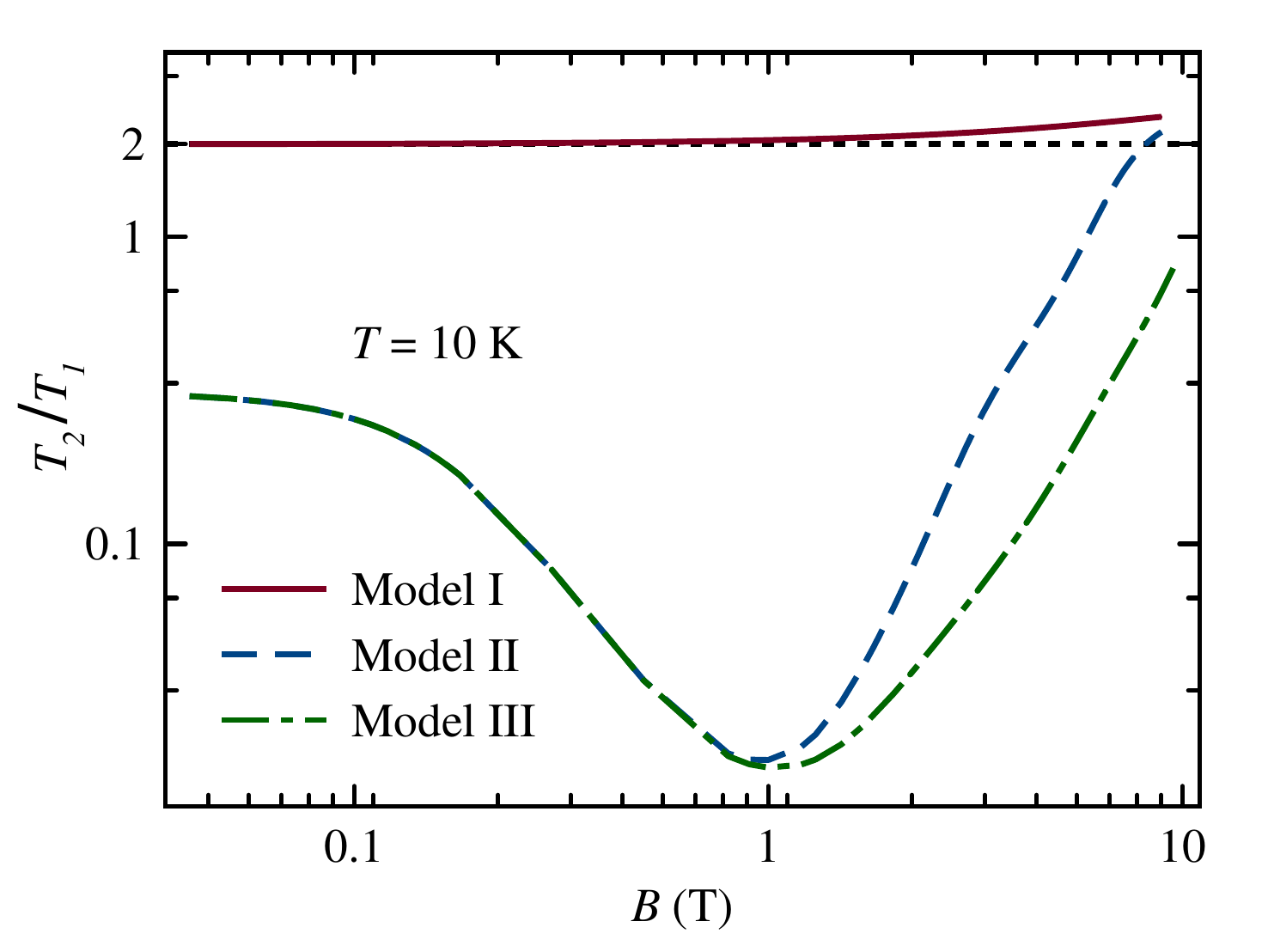}
\caption{$T_{2}/T_{1}$ ratio as a function of magnetic field predicted by the spectral density model I-III at $T=10 \,{\rm K}$, with model parameters given in Table \ref{tab:parameters}. The dashed horizontal line shows the pure relaxation limit.} 
\label{fig:ratio vs field}
\end{figure}

The Redfield equation in the eigenbasis of $\hat H_S$ describes population transfer between spin sublevels and evolution of spin coherences. The Redfield tensor in Eq. (\ref{eq:Redfield tensor}) includes off-diagonal non-secular contributions that couple populations and coherences, depending on the details of the spectral density \cite{Pollard1996,Wang2015,Farina2019}. The relative importance of population and dephasing depends on the beheviour of the spectral density with frequency and temperature \cite{deVega2017}. For secular open quantum systems, populations and coherences must evolve independently \cite{Cattaneo_2019}. In this secular case, the dephasing and relaxation times are related by $1/T_{2} = 1/2T_{1} +1/T^{\ast}_{2}$, where $T_2^*$ is the pure dephasing time. Since pure dephasing does not involve transitions between system eigenstates, it is only present if the spectral density has finite amplitude at zero frequency. Therefore, only the spin-bath spectral density $J^{\delta B}$ contributes to pure dephasing. The phonon bath spectral density scales as $J^{\delta g}\sim \omega$ in the low frequency limit, and thus can only contribute to $T_2$ via spin relaxation (e.g., $1/2T_1$) and non-secular terms.

For systems where only $T_1$ measurements are available to parametrize the total spectral density, the spin bath model parameters $a$, $c$, and $\gamma_{\rm pd}$ are less constrained. There can be magnetic fields and temperatures for which different bath models give the same prediction for $T_1$, but differ in the estimation of $T_2$. This is illustrated in Table \ref{tab:parameters} for VO(acac)$_2$, where the three sets of model parameters quantitatively agree with $T_1^{\rm exp}$ data over a broad range of magnetic fields, but predict $T_2$ times that can differ by an order magnitude.

Figure \ref{fig:ratio vs field} shows the predictions for the dephasing-to-relaxation ratio $T_{2}/T_{1}$  as a function of magnetic field for VO(acac)$_2$ spin qubits. The three sets of model bath parameters from Table \ref{tab:parameters} are compared. As expected, when the spin bath is neglected (Model I), the pure relaxation limit $T_{2} = 2T_{1}$ is obtained for a broad magnetic field interval. Non-secular deviations are found at higher fields ($B>3 \,{\rm T}$), which slightly increase $T_2$ beyond the pure relaxation limit. By including the contribution to dephasing from the spin bath spectral density (Models II-III), $T_2$ drops below $T_1$ at low fields, with a zero-field ratio $T_2/T_1$ given $\sqrt{A(0)}$ [see Eq. (\ref{eq:AofB model})]. The ratio further decreases with increasing magnetic field, mostly due to the increase of  $T_1$ up to the crossover field $B\approx 1 \,{\rm T}$. For magnetic fields beyond the crossover, the qubit spin dynamics becomes increasingly dominated by phonon-induced relaxation, making the $T_2/T_1$ tend towards the pure relaxation limit at high fields. The spin bath model with field-dependent fluctuation amplitude (Model III) has a slower approach to the spin-lattice relaxation limit, as in this case $T_2\sim 1/B^{2}$.

\begin{figure}[t]
\centering
\includegraphics[width=0.5\textwidth]{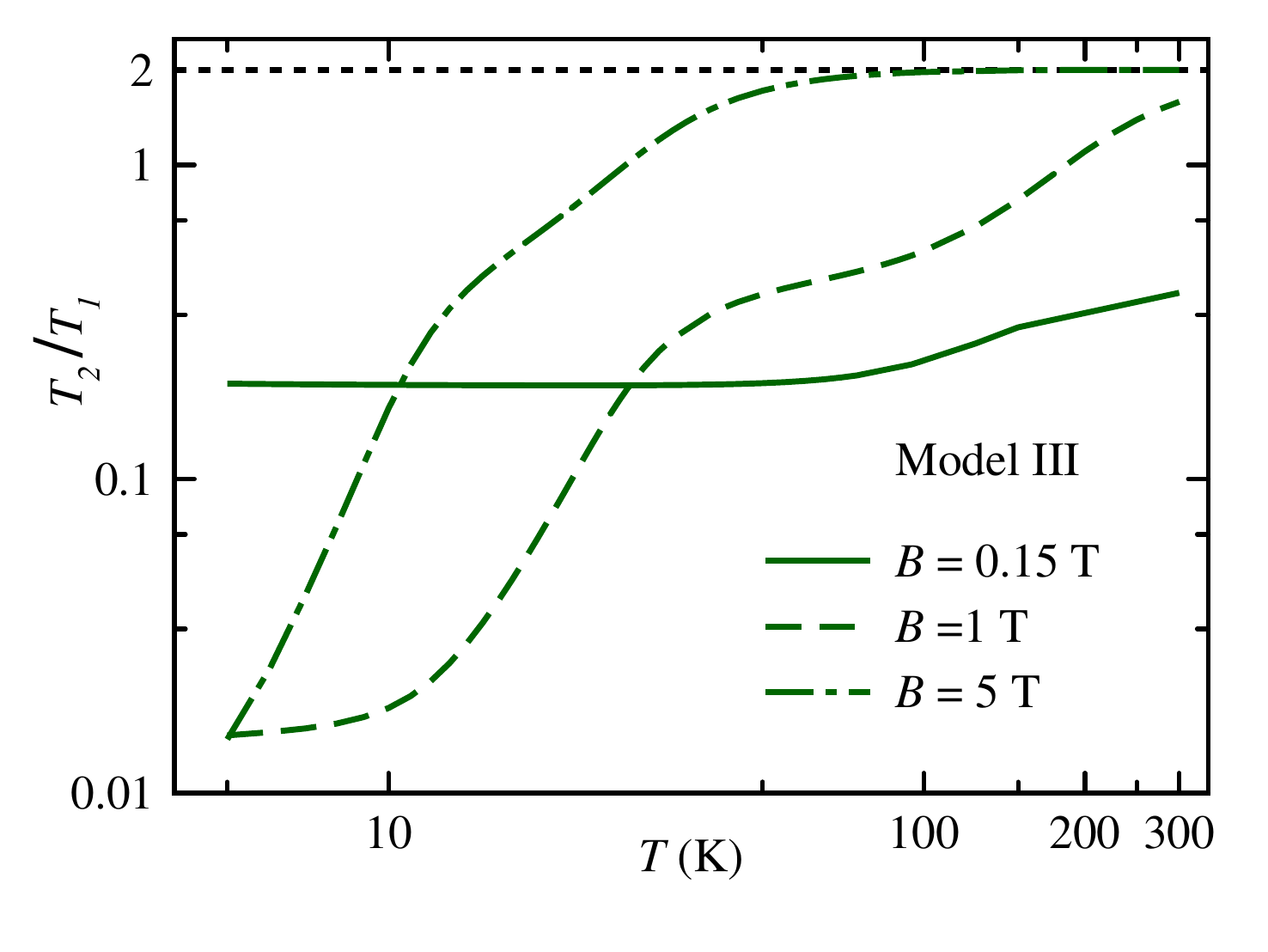}
\caption{$T_{2}/T_{1}$ ratio as a function of temperature predicted by the spectral density model III at different magnetic fields. Model parameters are given in Table \ref{tab:parameters}. The dashed horizontal line shows the pure relaxation limit.} 
\label{fig:ratio vs temperature}
\end{figure}   

Figure \ref{fig:ratio vs temperature} shows the temperature dependence of the ratio $T_{2}/T_{1}$ predicted by Model III for the spin bath fluctuations. We highlight this particular bath model because it shows the behaviors that could be described with this semi-empirical approach. For low magnetic fields ($B = 0.15$ T) the ratio $T_2/T_1$ is largely insensitive to temperature, with only a modest increase above $T\sim 100$ K due to the decrease of $T_1$. The predicted temperature dependence is however different at higher magnetic fields. For $B \gtrsim 5$ T, the ratio is an order of magnitude lower than the low-field result, but rapidly rises to reach the pure relaxation limit as the temperature is increased. Intermediate fields around $B\approx 1$ T give weaker increase of $T_2$ with temperature relative to higher magnetic fields. 

The temperature dependence of $T_2$ is directly related with the relative strength of the qubit coupling to the phonon and spin baths, characterized by the spectral densities $J^{\delta g}$ and $J^{\delta B}$, respectively. We only obtain $T_2<T_1$ over the entire range of temperatures ($5-300$ K) when the magnetic field noise is not negligible. In our model this implies that the qubit frequency (thus $B$-field) is below a threshold value, as in Fig. \ref{fig:ratio vs temperature}. Another possibility would be to introduce a temperature dependence to the zero-field fluctuation parameter, i.e., $a\rightarrow a_0T^\beta $, with a scaling parameter $\beta$ to be determined from experiments or {\it ab-initio} simulations.

\section{\label{sec:ConAndOut}Conclusions and outlook}

We introduced a semi-empirical Haken-Strobl method to construct Redfield quantum master equations that describe the population and coherence dynamics of solid-state molecular spin qubits in an external static magnetic field, as a function of temperature. In Haken-Strobl theory, the interaction of the molecular spin with lattice phonons is treated as a stochastic fluctuation of the gyromagnetic tensor and the interaction with other atomic spins in the crystal as fluctuations of the local magnetic field experienced by the qubit. Analytical expressions for the associated bath spectral autocorrelation functions can be parametrized using a small set of experimental measurements. We demonstrate the parametrization procedure using  experimental $T_1$ data from Ref. \cite{Tesi2016} for the vanadium(IV) spin qubit $\mathrm{VO(acac)_{2}}$ at different temperatures. Quantitative agreement is obtained for spin relaxation measurements at high magnetic fields, where spin-lattice relaxation dominates. Good qualitative agreement is found at lower magnetic fields, where spin-spin interactions become dominant. The model captures the experimentally observed optimal $T_1$ time as a function of magnetic field strength, which marks the crossover from a regime dominated by the spin-bath interaction at low magnetic fields, from the regime where qubit relaxation is dominated by spin-lattice interaction. 

The generalization ability of the proposed semi-empirical approach beyond the set of conditions used to parametrize the Redfield tensor can improve significantly by comparing the model prediction with  simultaneous measurements of the dephasing and relaxation times, $T_2$ and $T_1$, over a broad range of temperatures $T\sim 1-10^2$ K and magnetic fields $B\sim 0.1-10$ T. Complementary efforts could include the analysis of {\it ab-initio} dynamical results on the fluctuation of gyromagnetic tensor due to thermal lattice dynamics and the explicit modeling of the   hyperfine interaction of the molecular qubit with the spin-bath, which would allow for a microscopic treatment of the temperature dependence of the local magnetic field fluctuations. Such model improvements would enable the precise characterization of the qubit coherence time $T_2$ at room temperature, a key feature to optimize in the development of  quantum magnetometers based in solid-state molecular magnets \cite{Amdur2022}. 
 
\begin{acknowledgments}
KA is supported by ANID though Beca Doctorado 21220245. YC thanks the University of Notre Dame for financial support through startup funds. DA thanks ANID Fondecyt Regular 1210325 for support. FH  is supported by ANID through Fondecyt Regular 1221420, Millennium Science Initiative Program ICN17\_012. Work supported by the Air Force Office of Scientific Research under award number FA9550-22-1-0245.

\end{acknowledgments}

\bibliographystyle{apsrev4-1}
\bibliography{references}

\end{document}